\documentclass{ws-procs9x6}
\usepackage{epsfig} 
\newcommand{\quq}{\theta_{13}}
\newcommand{\nmt}{\nu_\mu \rightarrow \nu_\tau}
\newcommand{\nme}{\nu_\mu \rightarrow \nu_e}

\newcommand{\dmsq}{\Delta m^2}

\begin{document}

\title{Detectors for New Neutrino Experiments}

\author{Maury Goodman}

\address{Argonne National Laboratory\\Argonne IL 60439, USA\\
E-mail: maury.goodman@anl.gov}

\maketitle

\abstracts{
There has been great progress in understanding the neutrino sector
in the last few years.  One mixing angle has not yet been measured,
$\quq$.  
I will review detectors for
the current round of long-baseline neutrino oscillation 
experiments, for future long-baseline off-axis detectors, and for two
detector experiments at nuclear reactors.}

\section{Introduction}

In the last few years we have seen remarkable progress in understanding 
the neutrino.
Compelling evidence for the existence of neutrino mixing and 
oscillations has been 
presented in experiments using neutrinos from the 
atmosphere\cite{bib:sk1,bib:sim}, from the sun\cite{bib:sno},
from reactors\cite{bib:kamland} and from accelerators\cite{bib:k2k}.
Arguably, the most compelling task facing the neutrino community is
to measure a non-zero value of $\quq$.  If
this is non-zero, then future searches for CP violation and matter
effects are possible.  
There are two main ways to search for a non-zero value of
$\quq$.  They are new detectors in long-baseline neutrino beams, which
might be put about 2 degrees off the center of the beam,
 and
new reactor neutrino experiments with 2 detectors, one near and the other
about 2 km away.

\section{Current Long-Baseline Neutrino Experiments}
\par A current program of long-baseline neutrino experiments includes the
K2K experiment in Japan, which has been running since 1998, the NuMI/MINOS
program at Fermilab, which is expected to start in 2005, and the CNGS program
at CERN which will start in 2006.  Accelerator beams consist mostly of muon 
neutrinos ($\nu_\mu$) which are made when pions and Kaons decay.  The dominant
oscillation mode, which depends on the angle $\theta_{23}$ is
$\nmt$.  All three experiments expect to measure a change in the number and
distribution of $\nu_\mu$ as a result of this oscillation.   In addition,
CNGS hopes to measure $\nu_\tau$ appearance.  $\nu_\tau$'s would be present
in all three experiments, but the production of $\tau$'s in K2K and MINOS
is expected to be suppressed by kinematic thresholds.  All three experiments
will also have some sensitivity to $\quq$ by looking for $\nme$ oscillations
manifested by electron appearance.  This is because
\begin{equation}P(\nme) = \sin^2(2 \quq) \cos^2 (\theta_{23}) \sin^2 (\dmsq L/4E)
\end{equation}
Currently $\dmsq$ is near $2.0 \times 10^{-3} eV^2$ and $\theta_{23}$ near $45^o$
from atmospheric neutrino measurements, but $\quq$ is unknown, limited
by the Chooz reactor neutrino experiment as a function of $\dmsq$.

\par One long-baseline neutrino oscillation experiment is currently running,
K2K from the 12 GeV PS at KEK to Super-Kamiokande.  The
remarkable Super-Kamiokande experiment is a water Cerenkov detector, arranged
in a large stainless steel cylinder 37 m high by 34 m diameter.  It ran for five 
years with high (40\%) phototube coverage until it stopped for maintenance
in August 2001.   At the time, the tank was drained and some bad phototubes
were replaced.  While it was refilling, a tube imploded, causing a chain
reaction which destroyed more than half of the phototubes.  Since then
it has been rebuilt with approximately 20\% phototube coverage.  Extensive
tests were undertaken right after the accident to understand the mechanism.
It was determined that  a similar incident involving the propagation
of a shock wave could not take place if the 
phototubes were placed in a container with a small hole for water to
get in and out.  Thus the remaining phototubes were enclosed in  
newly designed vessels with a 13 mm acrylic front and a 5 mm fiberglass
back molded in a shape similar to the phototubes.  The
detector was rebuilt and filled in October 2002, and is once again running.
A beam from KEK is also running again.  There have been less than
100 fully contained events in K2K, which should double.
\par The MINOS collaboration has built a detector in the Soudan mine in
Minnesota to measure neutrinos produced by the NuMI beam at Fermilab.
MINOS has chosen a magnetic iron sampling calorimeter for both its near
and far detectors.  Both detectors consist of 2.54 cm steel absorber plates
and plastic scintillator planes.  The planes contain strips of extruded 
scintillator
which are 4.1 cm in width, and up to 8 meters long, with a wavelength 
shifting fiber carrying light to a Hamamatsu M16 phototube.  By reading
out both ends of the fiber, ionizing radiation is measured.

\par The CERN neutrino program is aiming a high energy beam
at two detectors in Italy's Gran Sasso Lab.
ICARUS is a very-high resolution liquid argon time projection chamber,
and OPERA is made from a lead-emulsion sandwich.  A 600 ton version
of ICARUS has been built and tested, with plans for
3000 tons.
OPERA will measure evidence for the production of $\tau$ decays
from $\nu_\tau$ charged current interactions produced after
oscillations by seeing evidence for a tau kink in photographic
emulsion.  They expect a signal of 18.3 $\times [\dmsq/(3.2 \times
10^{-3} eV^2]^2$ events with a background of 0.57 events in 2 years. 

\section{Future Off-axis Long-Baseline Neutrino Experiments}

\subsection{JPARC}
The Japanese Particle Research Center (JPARC) is a new 50 GeV
proton synchrotron being built in Tokai.  A neutrino
beam is being planned.  In a first phase the accelerator would operate
at 0.77 MW, but an upgrade to 4 MW 
is being considered.  The 22.5 kiloton Super-Kamiokande detector
is 295 km away, and the beam could be built to be simultaneously
a few degrees off axis to that experiment and to the proposed site
for a 1000 kiloton Hyper-Kamiokande detector in Tochibora.
With a 5 year run of JPARC, and the proposed 2 degree off-axis
beam, JPARC$\nu$ would be able to measure $\quq$ or set a limit
on $\sin^2(2\theta_{13})~<~0.006$ at 90\%~CL.  The Hyper-Kamiokande detector
would be similar in design to Super-K, using large 50 cm diameter
Hamamatsu phototubes.
\subsection{NuMI off-axis}

\par A proposal is being developed for an off-axis experiment using
the NuMI beam at Fermilab.\cite{bib:p929}  Any detector would be built near
the surface of the earth, about 10 km away from the center of the NuMI
beam.  The detector should have
a mass of 50 Kilotons, and be sensitive to 1 GeV electron showers.
The passive detector is planned to be 7 sheets of 2.5 cm particle board
between readout planes.
Active detector technologies being considered are resistive plate 
chambers and liquid and solid scintillator.  
\par To reduce costs from
the MINOS experience, the fibers in a scintillator detector would be read
out by Avalanche Photo Diodes (APDs), a low gain solid state detector with high
quantum efficiency built onto a chip.   The cost per channel in bare die form in the
quantities appropriate for this experiment (600,000) is \$2.70 per channel,
to be compared with a cost of about \$12 for similar quantities of multi-channel
photomultiplier tubes.  The APD quantum efficiency is 85\% in the wavelength
region of interest compared to 10\% for a PMT with bialkali photocathode.
\par The liquid scintillator design is for 14.4 m long multicell extrusions
of PVC, each containing 32 cells of width 3.75 cm.  The cells would be 3 cm
thick along the beam direction.  A looped fiber would be inserted in each cell
and an end-cap would be glued on one end, with a manifold/optical connector
assembly at the other.  There are no critical tolerances, such as positioning
of the fiber.  
\par One idea to standardize and stack large modules of detectors is 
to build them in standard shipping containers which are used throughout
the shipping industry on trucks, trains and internationally on large ships.
There is an ISO standard for these containers and as a result of the trade
imbalance, there is a large excess of them in the United States.

\section{Reactor Experiments}
\par
\ From the discovery of the neutrinos by Reines and Cowan\cite{bib:reines}
at Savannah River to the evidence for $\bar{\nu}_e$ disappearance at
KamLAND\cite{bib:kamland}, reactor neutrino experiments have studied
neutrinos in the same way -- observation of inverse beta decay with
scintillator detectors.  Since the signal from a reactor falls with
distance L as 1/L$^2$, as detectors have been moved further away from
reactors, it has become more important to reduce
backgrounds.  That can be done by putting experiments underground;
and experiments one kilometer or more away from reactors (Chooz, Palo Verde
and KamLAND) have been
underground.
\par The KamLAND experiment measured a 40\% disappearance of
$\bar{\nu}_e$ presumably associated with the 2nd term in Equation \ref{eq:p}:
\begin{equation}
\label{eq:p}
P(\bar{\nu}_e \rightarrow \bar{\nu}_e) \cong 
- \sin^2 2 \theta_{13} \sin^2(\Delta m^2_{atm} L/4E)  
- \cos^4 \quq \sin^2 2 \theta_{12} \sin^2(\Delta
m^2_{12} L/4E) + 1
\end{equation}
The Chooz and Palo Verde data put a limit on $\quq$ (through the first
term in Equation \ref{eq:p}) of $\sin^2 2 \theta_{13} < 0.1$.  
Those experiments could not have had greatly improved sensitivity
to $\quq$ because of uncertainties related to knowledge of the flux of 
neutrinos from the reactors.  They were designed to test
whether the atmospheric neutrino anomaly might have been due to
$\nme$ oscillations, and hence were searching for large mixing.
\par New experiments\cite{bib:kr2det,bib:kash} to look for non-zero values of $\quq$ would
need the following properties:
1)  two or more detectors to reduce uncertainties to the reactor flux
2) identical detectors to reduce systematic errors related to detector
acceptance,  3)
carefully controlled energy calibration,
4) low backgrounds and/or reactor-off data.  
\par
\begin{figure}
\begin{center}
         \mbox{\epsfig{figure=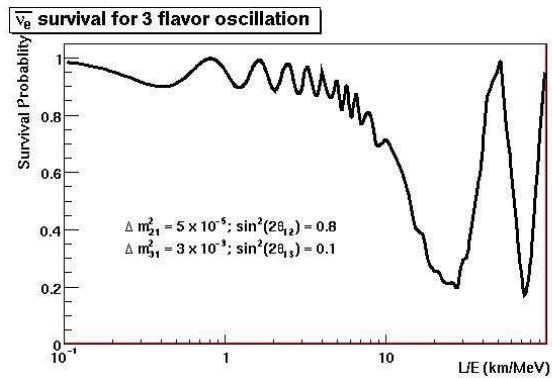,width=8.0cm}}
\caption{Probability of $\nu_e$ disappearance versus L/E for 
$\quq$ at its current upper limit}
\label{fig:P}
\end{center}
\end{figure}

 In Equation \ref{eq:p}, the values of $\theta_{12}$, $\Delta m^2_{12}$
and $\Delta m^2_{atmo}$ are approximately known.  In Figure \ref{fig:P},
the probability of $\bar{\nu}_e$ disappearance as a function of L/E
is plotted with  $\quq$ 
put at its maximum allowed value.  Note that
CP violation does not affect a disappearance experiment, and that
matter effects can be safely ignored.  The
large variation in P for L/E$>$ 10 km/MeV is the effect seen by
KamLAND and solar $\nu$ experiments.  The much smaller deviations from
unity for L/E $<$ 1 km/MeV are the goal for an accurate new reactor
experiment.  The detectors are planned to be large volumes of
liquid scintillator 25-100 tons, similar to the 10 ton Chooz experiment, but
smaller than the 1100 ton KamLAND detector.

\end{document}